\documentclass[nanomaterials,article,accept,pdftex,moreauthors]{Definitions/mdpi}
\firstpage{1} 
\pubvolume{1}
\issuenum{1}
\articlenumber{0}
\pubyear{2023}
\copyrightyear{2023}
\externaleditor{Academic Editor: Zhidong Zhan}
\datereceived{11 January 2023} 
\daterevised{25 January 2023} 
\dateaccepted{29 January 2023} 
\datepublished{ } 
\hreflink{https://doi.org/} 

\usepackage[normalem]{ulem}

\Title{Effects of the Spatial Extension of the Edge Channels on the Interference Pattern of a Helical Josephson Junction}

\TitleCitation{Effects of the Spatial Extension of the Edge Channels on the Interference Pattern of a Helical Josephson Junction}

\Author{Lucia Vigliotti $^{1,}$*\orcidA{}, Alessio Calzona $^{2}$\orcidB{}, Niccolò Traverso Ziani $^{1,3}$\orcidC{}, F. Sebastian Bergeret $^{4,5}$\orcidD{}, Maura Sassetti $^{1,3}$ \mbox{and Bj\"orn Trauzettel $^{2,6}$\orcidF{}}}

\AuthorNames{Lucia Vigliotti, Alessio Calzona, Niccolò Traverso Ziani, F. Sebastian Bergeret, Maura Sassetti and Bj\"orn Trauzettel}

\AuthorCitation{Vigliotti, L.; Calzona, A.; Traverso Ziani, N.; Bergeret, F.S.; Sassetti, M.; Trauzettel, B.}

\address{%
$^{1}$ \quad Dipartimento di Fisica, Università degli Studi di Genova, Via Dodecaneso 33, 16146 Genova, Italy; traversoziani@fisica.unige.it (N.T.Z.); sassetti@fisica.unige.it (M.S.)\\
$^{2}$ \quad Institute for Theoretical Physics and Astrophysics, University of W\"urzburg, D-97074 W\"urzburg, Germany; alessio.calzona@physik.uni-wuerzburg.de (A.C.); trauzettel@physik.uni-wuerzburg.de (B.T.)\\
$^{3}$ \quad CNR-SPIN, Via Dodecaneso 33, 16146 Genova, Italy\\
$^{4}$ \quad Centro de F\'isica de Materiales (CFM-MPC), Centro Mixto CSIC-UPV/EHU, \mbox{E-20018 Donostia-San Sebasti\'an, Spain}; fs.bergeret@csic.es\\
$^{5}$ \quad Donostia International Physics Center (DIPC), E-20018 Donostia–San Sebasti\'an, Spain\\
$^{6}$ \quad W\"urzburg-Dresden Cluster of Excellence ct.qmat, Germany}

\corres{Correspondence: lucia.vigliotti@edu.unige.it}

\abstract{Josephson junctions (JJs) in the presence of a magnetic field exhibit qualitatively different interference patterns depending on the spatial distribution of the supercurrent through the junction. In  JJs based on two-dimensional topological insulators (2DTIs), the electrons/holes forming a Cooper pair (CP) can either propagate along the  same edge or be split into the two edges. The former leads to a SQUID-like interference pattern, with the superconducting flux quantum $\phi_0$ (where $\phi_0=h/2e$) as a fundamental period. If CPs' splitting is additionally included, the resultant periodicity doubles.  Since the edge states are typically considered to be strongly localized, the critical current does not decay as a function of the magnetic field. The present paper goes beyond this approach and inspects a topological JJ in the tunneling regime featuring extended edge states. It is here considered the possibility that the two electrons of a CP propagate and explore the junction independently over length scales comparable to the superconducting coherence length. As a consequence of the spatial extension, a decaying pattern with different possible periods is obtained. In particular, it is shown that, if crossed Andreev reflections (CARs) are dominant and the edge states overlap, the resulting interference pattern features oscillations whose periodicity approaches $2\phi_0$.}


\begin{document}


\section{Introduction} \label{sec:introduction}

Topological phases of quantum systems have been at the forefront of research in condensed matter over the last two decades \cite{hasan,qi}. One of these phases takes place in quantum spin Hall (QSH) insulators, which are two-dimensional topological insulators (2DTIs) hosting topologically protected and counter-propagating helical edge states on their \mbox{boundary \cite{dolcetto,bernevig,strunz,konig,knez,reis,wu,liu}}. The interplay of superconductivity and the QSH effect has been widely studied in view of applications in spintronics and in (topological) quantum computation \cite{hart,pribiag}. To this end, topological Josephson junctions (JJs) appear as fundamental building blocks \cite{deacon,dolcini}. In a topological JJ, two superconducting electrodes are connected through the helical edge state channels of the QSH insulator. If the junction is pierced by a magnetic flux, it realizes a superconducting quantum interference setup \cite{hart,pribiag,bocquillon}. The interference pattern, namely the flux dependence of the critical current, characterizes JJs. Despite many theoretical studies on the interference patterns, there are still open questions, particularly when it comes to comparison with experiments \cite{ke,suominen,kurter,borcsok}.

\begin{figure}[H]
		\includegraphics[width=10cm]{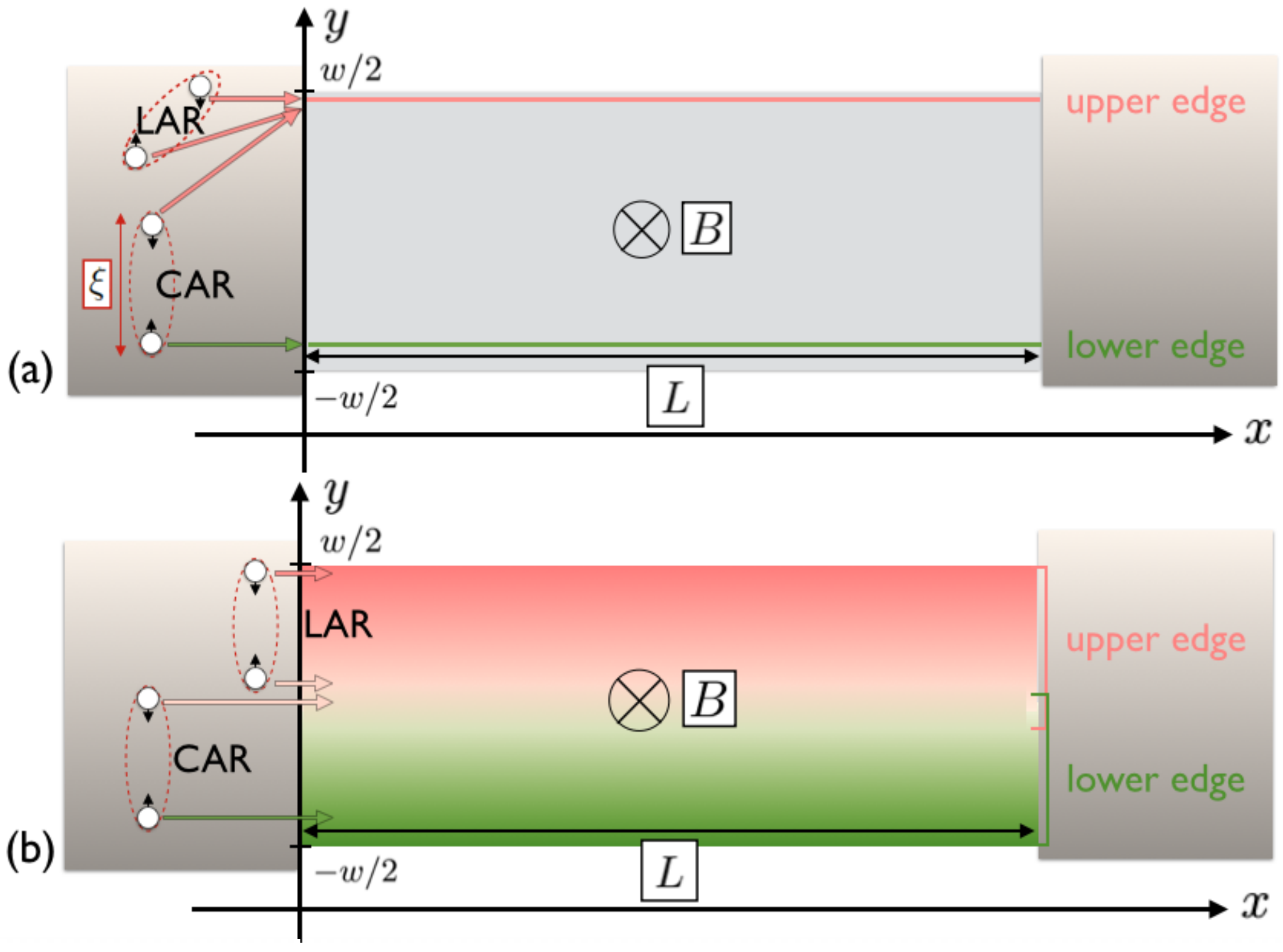}
		\caption{Panel (\textbf{a}) shows a two-dimensional topological insulator (2DTI) of length $L$ and width $w$ laterally tunnel-coupled to two superconducting electrodes (in grey). Then, a magnetic field B is applied perpendicularly to the junction. The pink and green line represent the edge states on the boundaries of the 2DTI which are not proximitized. Each boundary hosts two counter-propagating channels with identical profiles. For clarity, only one colored shape per boundary is shown. The electrons forming a Cooper pair (CP) can be injected into a same edge via a local Andreev reflection (LAR) or into opposite edges via a crossed Andreev reflection (CAR). The CP splitting of the latter is allowed only if the superconducting coherence length $\xi$, which is effectively the size of the CP, is larger or comparable with the width $w$. Panel (\textbf{b}) shows the same sample for the case of extended edge states, which allows different trajectories for electrons injected in LAR and a wider range of possibilities for CAR. For clarity, both panels show LAR processes involving only the upper edge.}
		\label{fig:wide}
\end{figure}

Many established models for JJs usually assume a local transmission of the Cooper pairs (CPs), {i.e.,} the same trajectory for both electrons \cite{tinkham,barone}. A non-local transmission is also considered in the framework of edge transport via CPs' splitting over opposite edges \cite{sato,devries,baxevanis,recher,blasi}; this is allowed over length scales comparable with the superconducting coherence length $\xi=\hbar v_F/\Delta$, with $v_F$ as the Fermi velocity and $\Delta$ as the superconducting gap, but usually discussed in the case of narrow edge states (see the upper panel of Figure~\ref{fig:wide}). Specifically, strongly localized edge states give rise to a sinusoidal double-slit pattern, similar to a SQUID pattern, with no decay and a period $\phi_0=h/2e$. However, the presence of interference oscillations with a doubled periodicity has been theoretically predicted \cite{baxevanis,klapwijk,galambos} and experimentally observed in different setups \cite{klapwijk,harada}, including 2DTI-based JJs. In this case, the origin of this doubling relies on the CAR processes mentioned above: a non-local transmission of electrons (whose charge quantum is $e$ versus the CP's charge quantum of $2e$) takes \mbox{place \cite{baxevanis,galambos}}. Depending on the amount of CPs' splitting, the resultant pattern features either alternating lobes with different heights or a weak cosine modulation around a constant value. On top of that, further single-electron effects leading to anomalous periodicities such as back-scattering \cite{mironov} or forward-scattering \cite{vigliottiNJ} have been assessed. Lastly, it is worth recalling that a moderate spatial extent of the edge states affects the interference pattern with an overall decay in the magnetic field \cite{hart}.

Although the scenario of extended edge states might be experimentally relevant, a theoretical model is still lacking. This is addressed in this article by means of a heuristic approach. An edge state with finite spatial extension can host different trajectories for the two electrons forming the CP, provided that they are not further away from each other than $\xi$, injected either into a same edge (local Andreev reflection, LAR) or into different edges (crossed Andreev reflection, CAR) (see Figure~\ref{fig:wide}). The wider the edges, the more pronounced will be the consequences on the interference pattern, which is highly sensitive to the electrons' path.

In this work, the combined effect of broadened edge channels, possibly overlapping, and the presence of CAR is explored. This introduces new options for the injection process, which are absent in the case of narrow edges and which enrich the possibilities of interference patterns. Differently from previous approaches assessing 2DTI junctions, a fast side lobe decay and different oscillation periods are obtained. Within this wide phenomenology, it is interesting to discuss whether CAR processes can bring along a doubled periodicity as in the case of localized edge states. It is found that the answer is affirmative and the regime to observe such periodicity is identified, finding that it requires a prevalence of CAR over LAR.

The main findings of this work are: the derivation of an expression that allows for the computation of supercurrents in the experimentally relevant scenario of topological Josephson junctions featuring edge states with finite spatial extent; and the introduction of a new way of taking into account the non-local character of CPs.

To simplify the problem, the following assumptions are made throughout the text: the interfaces between the superconductors and the non-superconducting region are assumed to be low transparent, leading to a sinusoidal current-phase relation; the two edge states are assumed to be symmetric in shape; and trajectories other than horizontal ones and inter-edge tunneling are neglected.

The article is structured as follows: in Section~\ref{sec:transport}, the way of  calculating  the Josephson current through a junction is reviewed by introducing the gauge-invariant phase. Both local and non-local transfers of CPs are addressed. In Section~\ref{sec:model}, the model and approach to determine the supercurrent are introduced. Sections~\ref{sec:results} and~\ref{sec:discussions} are devoted to the presentation of the main results and a more general discussion, respectively. Finally, in Section~\ref{sec:conclusions}, conclusions are drawn.


\section{Local and Non-Local Transport of Cooper Pairs} \label{sec:transport}

Let us consider a two-dimensional JJ of length $L$ and width $w$ as in Figure~\ref{fig:wide}. The intermediate region is tunnel-coupled to two superconductors on either side and for now, it is not needed to further specify its properties. A magnetic field is applied perpendicularly to the plane, $\mathbf{B}=B\mathbf{e_z}$. It is here assumed that the field is screened from the superconducting electrodes, and the gauge $\mathbf{A}=-By\mathbf{e_x}$, where $\mathbf{A}$ is the vector potential, is chosen.

 For the evaluation of the supercurrent, it is convenient to introduce the gauge-invariant phase difference, $\delta\theta=\delta\varphi-(2\pi/\phi_0)\int \mathbf{A}\cdot d\mathbf{r}$, with $\delta\varphi$ the superconducting phase \mbox{difference \cite{tinkham,tkachov}}. The gauge-invariant phase picked by a CP being transmitted across the junction along a horizontal (ballistic) path $y$, with $-w/2<y<w/2$, is then  given by
 
\begin{equation}
	\delta\theta(y)=(\varphi_r-\varphi_l)+\frac{2\pi\phi}{\phi_0}\frac{y}{w}.\label{eqn:phase}
\end{equation}
Here, $\varphi_{r/l}$ labels the right/left superconducting phase, and $\phi=BLw$. The second term in Equation~\eqref{eqn:phase} stems for the Aharonov--Bohm contribution, which for a single electron reads as $\delta\theta^{AB}(y)=\frac{\pi\phi}{\phi_0}\frac{y}{w}$.

Concerning the computation of supercurrents, a standard approach is the Dynes and Fulton description \cite{dynes}, which holds in the tunneling regime (low-transparency interfaces) between the superconductors and their link under the assumption of the local nature of the supercurrent, flowing perpendicularly to the superconducting contacts. This means that the supercurrent density only depends on the $y$ coordinate while the current flows in the $x$ direction. In this case, for the junction just introduced, the total current is given by

\begin{equation}
    I(\phi,\varphi_r-\varphi_l)=\int_{-w/2}^{w/2}dy\,j(y)\sin{\left[(\varphi_r-\varphi_l)+\frac{2\pi\phi}{\phi_0}\frac{y}{w}\right]},\label{eqn:dynesfulton}
\end{equation}
with $j(y)$ being the current density profile of the JJ. The total current therefore results from a weighted integration over sinusoidal current-phase-relations (stemming from the tunneling regime). Maximizing with respect to $(\varphi_r-\varphi_l)$ and getting the absolute value, one obtains the critical current or interference pattern $I_C(\phi)$. This procedure recovers well-known examples of interference patterns \cite{barone}: for an uniform current distribution $j(y)=I_C/w$ ($I_C$ being a constant), it reproduces the Fraunhofer pattern, $I_C(\phi)/I_C(\phi=0)=\left|\sin{(\pi\phi/\phi_0)}/(\pi\phi/\phi_0)\right|$; if there is only edge transport, and the edge channels are assumed to be extremely narrow, $j(y)\propto[\delta(y-w/2)+\delta(y+w/2)]/2$, and one gets the SQUID pattern, $I_C(\phi)/I_C(\phi=0)=\left|\cos{(\pi\phi/\phi_0)}\right|$.

Non-local transmission has been previously addressed in different realizations of \mbox{JJs \cite{park,klapwijk,barzykin,ledermann,sheehy}}. This work focuses on JJs featuring edge states, usually modeled as strongly localized. In these setups, a sample's width $w$ comparable with the superconducting coherence length $\xi$ allows an effective splitting of the CP via CAR. In this case, the Aharonov--Bohm phases acquired by the electrons propagating on opposite edges cancel, resulting in a flux-independent process. This leads to the $2\phi_0$-periodic even--odd effect in SQUID-like patterns, which has been experimentally observed \cite{pribiag,devries,devries2} and theoretically \mbox{addressed \cite{baxevanis,blasi,galambos}} in several works. Such phenomenology is shared by helical and non-helical edge channels, though remarkable qualitative differences emerge in response to variations of the parameters \cite{galambos}. Besides the even--odd effect, it has been discussed how inter-channel scattering events give rise to anomalous flux dependencies leading, for instance, to multi-periodic magnetic oscillations \cite{mironov} or to a further doubling of the period up to $4\phi_0$ \cite{vigliottiNJ}.

In the following, it is discussed how the current can be calculated in two-dimensional systems with extended edge states. Different interference patterns that depend on the extension of the edge states and on the width of the junction are found. The finite extension of the edge states leads to a Fraunhofer-like interference pattern, with a main central lobe and decaying side lobes. In particular, it is shown that, if CARs are dominant and the edge states overlap, the resulting periodicity approaches $2\phi_0$.


\section{Model for Extended Edge Channels} \label{sec:model}
The system under consideration is a junction as the one depicted in the lower panel of Figure~\ref{fig:wide}, consisting of a two-dimensional JJ where the weak link is a topological insulator sample of length $L$ and width $w$. This region is tunnel-coupled to the right and left superconductors. As previously, the phase of the right/left superconductor is denoted as $\varphi_{r/l}$. Due to the proximity effect, in the superconducting parts, the edge states are gapped out. In the center region, the edge states are helical. In Figure~\ref{fig:wide}, each boundary hosts two counter-propagating channels with identical profiles. For clarity, only one colored shape per boundary is shown.

\begin{figure}[H]
	\includegraphics[width=5cm]{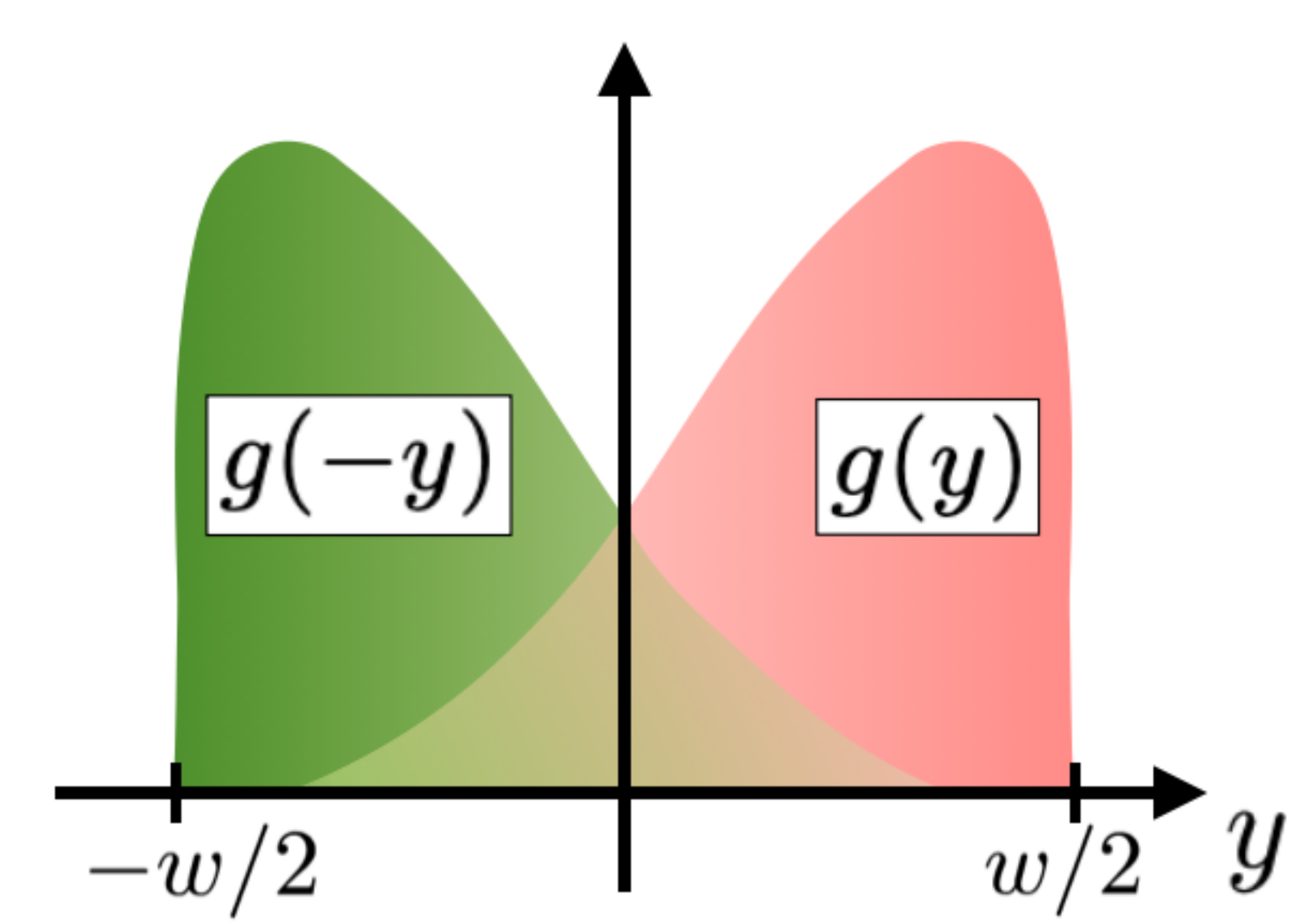}
	\caption{The colored broadened shapes represent the edges' profiles: $g(y)$ for the upper edge (pink) and $g(-y)$ for the lower edge (green). They are therefore symmetric around $y=0$ and overlap to some extent.}
	\label{fig:density}
\end{figure}

Following the line of reasoning in the previous section, it is possible to write a phenomenological expression for the supercurrent that generalizes Equation~\eqref{eqn:dynesfulton} with two different coordinates for the two electrons:

\begin{align}
	I(\phi,\varphi_r-\varphi_l)&=\int_{-w/2}^{w/2}dy_{\uparrow}\,dy_{\downarrow}\,j(y_{\uparrow},y_{\downarrow})\sin{\left(\varphi_r-\varphi_l+\frac{\pi\phi}{\phi_0}\frac{(y_{\uparrow}+y_{\downarrow})}{w}\right)}\nonumber\\
    &=\text{Im}\left[e^{i(\varphi_r-\varphi_l)}\int_{-w/2}^{w/2}dy_{\uparrow}\,dy_{\downarrow}\,j(y_{\uparrow},y_{\downarrow})e^{i\frac{\pi\phi}{\phi_0}\frac{y_{\uparrow}}{w}}e^{i\frac{\pi\phi}{\phi_0}\frac{y_{\downarrow}}{w}}\right],\label{eqn:ourformula}
\end{align}
where the fundamental ingredient is $j(y_{\uparrow},y_{\downarrow})$, the weight function for the supercurrent, and $y_{\uparrow},\,y_{\downarrow}$ label the horizontal trajectories of the two electrons of the CP, with $\uparrow/\downarrow$ denoting the spin projection. For now, neither diagonal trajectories nor any inter-edge tunneling are included. The function $j(y_{\uparrow},y_{\downarrow})$ parametrizes how each specific path contributes to the total supercurrent and encodes physical properties of the normal region, such as the supercurrent density profile, the number of transport channels, and the helical nature of the junction. If the size of the CP is comparable with the junction's width, the CP can be split into the two edges. Since broadened edge states are considered here, it is assumed that the CP can also be split into different trajectories within the same edge. To do so, an overall constraint function to take into account the CP's extent is included. The ansatz is hence the following

\begin{equation}
    j(y_{\uparrow},y_{\downarrow})=e^{-|y_{\uparrow}-y_{\downarrow}|/\xi}[\underbrace{sg(y_{\uparrow})g(y_{\downarrow})+sg(-y_{\uparrow})g(-y_{\downarrow})}_{LAR}+\underbrace{g(-y_{\uparrow})g(y_{\downarrow})}_{CAR}],\label{eqn:densityexpr}
\end{equation}
where $g(\pm y)$ describes the spatial extension of the upper/lower edge states, which are assumed to be symmetric around $y=0$ (see Figure~\ref{fig:density} for a schematic view). Since $j$ is a probability density, one can argue that $g(y)\equiv|\psi_l(-y)|=|\psi_u(y)|$, where $\psi_{u/l}(y)$ is the wavefunction of the upper/lower edge state. Our approach allows us to identify the CAR and LAR processes generalized to the case of extended edge states, as marked in Equation~\eqref{eqn:densityexpr}. There are two parameters to be discussed in the following: the coherence length $\xi$ and the ratio of the amplitudes of LAR and CAR processes, denoted by $s$. Indeed, due to helicity, LAR and CAR are clearly different processes. Since spin-flips are not considered, in the LAR case, spin-up and spin-down electrons have opposite directions of propagation. By contrast, in the CAR case, they are either right-movers or \mbox{left-movers \cite{galambos,vigliottiNJ}}.

Equations~\eqref{eqn:ourformula} and \eqref{eqn:densityexpr} show two main features: (1) the electrons can tunnel into the same edge but at different positions; (2) the electrons can tunnel into different edges acquiring Aharonov--Bohm phases that do not cancel each other out. The latter implies the unconventional possibility of flux-dependent CAR processes.
It is possible to check some limiting cases of Equations~\eqref{eqn:ourformula} and \eqref{eqn:densityexpr}. Firstly, as to LAR processes, notice that they recover the Dynes and Fulton approach for $\xi\ll w$ \cite{dynes}. One can rewrite $e^{-|y_{\uparrow}-y_{\downarrow}|/\xi}=e^{-\frac{|y_{\uparrow}-y_{\downarrow}|}{w}\frac{w}{\xi}}$, where the first fraction takes values between 0 and 1. Then $e^{-|y_{\uparrow}-y_{\downarrow}|/\xi}\overset{\xi\ll w}{\longrightarrow}0$, and the current density vanishes unless $y_{\uparrow}=y_{\downarrow}\equiv y$. In this case
    
\begin{equation}
    j(y_{\uparrow},y_{\downarrow})=j(y)\propto|\psi_u(y)|^2+|\psi_l(y)|^2,
\end{equation}
and the supercurrent recovers the form
    
\begin{equation}
    I(\phi,\varphi_r-\varphi_l)\propto\,\text{Im}\bigg[e^{i(\varphi_r-\varphi_l)}\int_{-w/2}^{w/2}dy\,\left(|\psi_u(y)|^2+|\psi_l(y)|^2\right)e^{i\frac{2\pi\phi}{\phi_0}\frac{y}{w}}\bigg],
\end{equation}
which is the Dynes and Fulton description in Equation~\eqref{eqn:dynesfulton}. On the other hand, if $\xi\gg w$ $e^{-|y_{\uparrow}-y_{\downarrow}|/\xi}\overset{\xi\gg w}{\longrightarrow}1$ and
    
\begin{equation}
    j(y_{\uparrow},y_{\downarrow})\propto |\psi_u(y_{\uparrow})||\psi_u(y_{\downarrow})|+|\psi_l(y_{\uparrow})||\psi_l(y_{\downarrow})|.
\end{equation}
The integrals over $y_{\uparrow}$ and $y_{\downarrow}$ factorize
        
\begin{align}
	I(\phi,\varphi_r-\varphi_l)\propto\,\text{Im}\bigg[e^{i(\varphi_r-\varphi_l)}\bigg(&\int_{-w/2}^{w/2}dy_{\uparrow}\,|\psi_u(y_{\uparrow})|e^{i\frac{\pi\phi}{\phi_0}\frac{y_{\uparrow}}{w}}\int_{-w/2}^{w/2}dy_{\downarrow}\,|\psi_u(y_{\downarrow})|e^{i\frac{\pi\phi}{\phi_0}\frac{y_{\downarrow}}{w}}+\nonumber\\
    &\int_{-w/2}^{w/2}dy_{\uparrow}\,|\psi_l(y_{\uparrow})|e^{i\frac{\pi\phi}{\phi_0}\frac{y_{\uparrow}}{w}}\int_{-w/2}^{w/2}dy_{\downarrow}\,|\psi_l(y_{\downarrow})|e^{i\frac{\pi\phi}{\phi_0}\frac{y_{\downarrow}}{w}}\bigg)\bigg],
\end{align}
corresponding to completely independent trajectories.

Regarding CAR, if the conduction can only happen on narrow edges (such as in the upper panel of Figure~\ref{fig:wide}), then $|\psi_{u/l}(y)|\propto\delta(y\mp w/2)$, which results in a flux-independent contribution to the critical current, as expected.

The dependence of $s$ on temperature, bias, or length of the junction is not specified. Instead, it is treated as a phenomenological parameter. The next aim of this work is to identify a parameter regime in which the interference pattern is $2\phi_0$-periodic. Indeed, as the doubled periodicity is a widely studied signature, it is interesting to investigate new mechanisms that can give rise to it. It has been discussed that it usually emerges in the presence of Cooper pair splitting, which is a main feature of our description of broadened edge states. It is therefore expected to arise in our system. It turns out that, in our model, CAR-dominated transport is required to obtain this unusual periodicity of the maximal critical current. It will therefore be assumed that $s<1$ from now on. (Notice that one of the two CAR contributions should be proportional to $s^2$. Since $s<1$, it will be neglected, and only the first order in $s$ will be included.) Notably, it has been experimentally revealed in InSb JJs \cite{devries} that CAR processes are larger than expected and can even exceed LAR. Indeed, an entirely $2\phi_0$-periodic pattern, in combination with an enhanced conduction at both edges, was measured. Such $2\phi_0$ periodicity can result from the flux-independent supercurrent due to CAR interfering with the standard $\phi_0$-periodic SQUID current. However, if LAR dominates over CAR, a $\phi_0$ oscillation should be simultaneously present. Not being the case, it was concluded that the CAR amplitude was larger than the LAR one. It is interesting to identify rather general conditions under which CAR processes are more important than LAR processes, but this analysis goes beyond the scope of the present work. 

So far, a formula has been constructed that generalizes the computation of a supercurrent given the current density to the case of extended edges and shown that it recovers the expected limiting cases. In the next section, it is shown that, in an appropriate parameter range and for wide edge states, our model features an interference pattern approaching a $2\phi_0$ Fraunhofer pattern.


\section{Main Results} \label{sec:results}

Here the interference pattern of the JJ is analyzed, discussing the role of the edges' profile $g(y)$ and the two parameters $\xi$ and $s$. Given Equation~\eqref{eqn:ourformula}, the pattern reads

\begin{equation}
	I_C(\phi)=\left|\int_{-w/2}^{w/2}dy_{\uparrow}\,dy_{\downarrow}\,j(y_{\uparrow},y_{\downarrow})e^{i\frac{\pi\phi}{\phi_0}\frac{y_{\uparrow}}{w}}e^{i\frac{\pi\phi}{\phi_0}\frac{y_{\downarrow}}{w}}\right|,\label{eqn:pattern}
\end{equation}
with $j(y_{\uparrow},y_{\downarrow})$ from Equation~\eqref{eqn:densityexpr}.

\begin{figure}[H]
	\includegraphics[width=.99\textwidth]{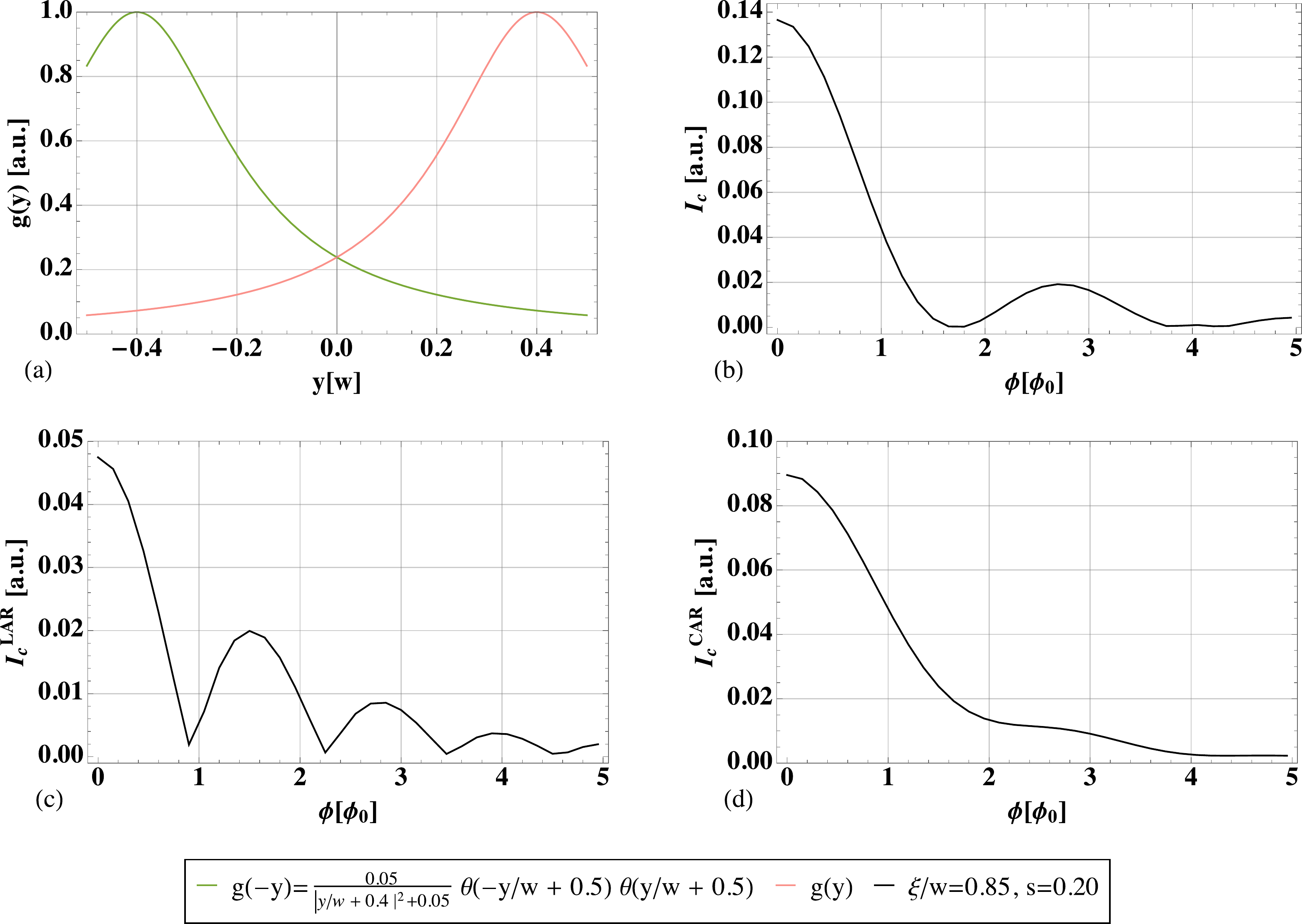}
	\caption{Resultant interference pattern $I_C(\phi)$ (panel (\textbf{b})) and the separated contributions of LARs (panel (\textbf{c})) and CARs (panel (\textbf{d})) for the edge profile in panel (\textbf{a}), $\xi/w=0.85$ and $s=0.2$.} 
	\label{fig:best}
\end{figure}

\begin{figure}[H]
	\includegraphics[width=0.5\textwidth]{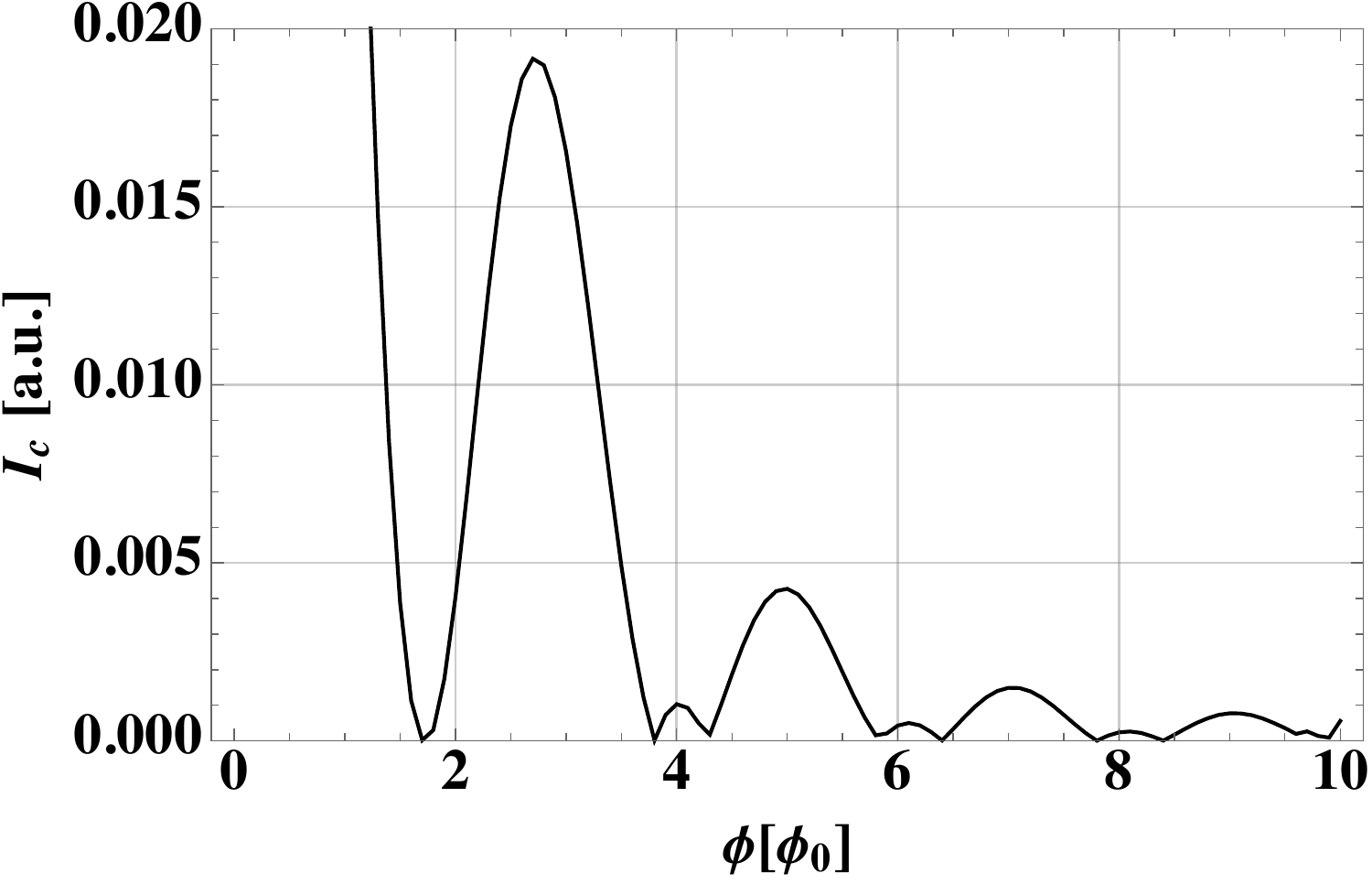}
	\caption{Resultant interference pattern $I_C(\phi)$ in Figure~\ref{fig:best}b for a larger interval of flux quanta.}
	\label{fig:zoom}
\end{figure}

Figure~\ref{fig:best} illustrates our results. It is assumed, as in the edge profile depicted in panel (a), $\xi/w=0.85$ and $s=0.2$. In Figure~\ref{fig:best}b, the total interference pattern is shown: it exhibits minima approaching multiples of $2\phi_0$ and a fast decay. In panels (c)-(d), the LAR contribution and the CAR term ($s=0$) alone, respectively, are plotted in order to point out the essential interplay of the two processes. On the one hand, the LAR pattern qualitatively resembles a standard Fraunhofer pattern, although its minima are shifted away from $\phi_0$ multiples as a consequence of the spatial extent of the edges. On the other hand, CAR processes feature a strong decay with a mild $2\phi_0$ modulation on top. The $2\phi_0$ oscillatory behavior in Figure~\ref{fig:best}b results from the interaction of these two terms. The interference patterns in Figure~\ref{fig:best} are shown for a limited number of flux quanta which, however, allow us to appreciate that the minima in $\phi=\phi_0,\,3\phi_0,\,5\phi_0$, which would be expected for a standard Fraunhofer-like pattern, are not visible. On the contrary, those in $\phi=2\phi_0,\,4\phi_0$ persist. For the sake of completeness, Figure~\ref{fig:zoom} shows the plot in Figure~\ref{fig:best}b for a larger interval, confirming this trend. Notice that for $\phi=4\phi_0,\,6\phi_0$ and also $\phi=8\phi_0$, the interference pattern does not completely vanish but presents very low peaks. These are reminiscent of the peak structure of the LAR contribution in Figure~\ref{fig:best}c, which is dominant over the CAR one for large values of $\phi$ due to its slower decay.

In the next section, the robustness of the effect is discussed by providing plots of the interference pattern for different values of the parameters. From such analysis, the optimal parameter range for the doubled minima periodicity is inferred. It is summarized as follows.
A high coherence length $\xi$ ($\xi\gtrsim w$) is necessary because short values of $\xi/w$ suppress the occurrence of CARs. This first requirement depends on the choice of the superconductors and on the sample width, and it is not hard to fulfill. The ratio $s$ has to be low (at least $s<1/2$), which means that CARs are dominant over LARs. A significant overlap of the edge states is needed. Indeed, it can be shown that, if the edge states do not overlap, the full interference pattern starts to exhibit the features expected for perfectly localized edges: it approaches a SQUID-like pattern with the additional even--odd effect, which is overall $2\phi_0$-periodic but not decaying.

\section{General Discussion} \label{sec:discussions}
A more general discussion is provided here, commenting on the interference pattern obtained for a wider range of parameters. This allows us to substantiate the optimal ranges stated in the main text.
 
\begin{figure}[H]
	\centering
	\includegraphics[width=\textwidth]{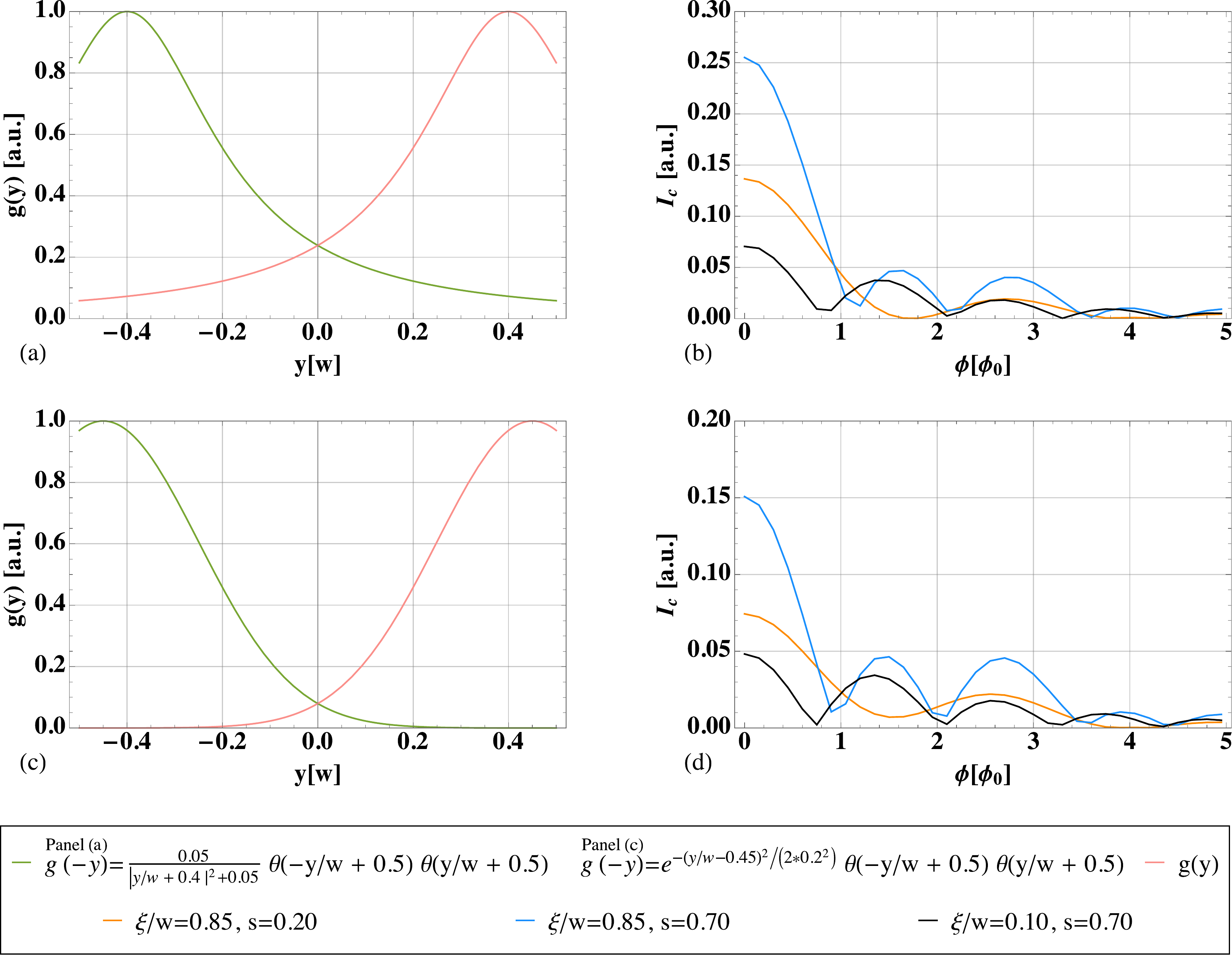}
	\caption{Flux-dependence of the critical supercurrent considering different values of coherence length, a more or less prevalent role played by LARs and CARs (represented by the parameter $s$), and different profiles for the edge states. In each row, the first panel (\textbf{a},\textbf{c}) shows profile $g(y)$ and the symmetric $g(-y)$. In the second column (panels \textbf{b},\textbf{d}), the full interference pattern, arising from both LARs and CARs, is plotted. Different colors are associated with different values of $\xi$ and $s$; see the plot legend.} 
	\label{fig:results}
\end{figure}

In Figure~\ref{fig:results}, two different shapes for the edge states are taken into consideration and plotted in the first column (the upper edge in pink, the lower one in green). They are both peaked at the opposite ends of the junction, around $y=\pm w/2$, but feature a decreasing overlap from the top row to the bottom row. In the second column, the full interference pattern arising from both LAR and CAR is plotted. Each colored line corresponds to a different combination of $\xi/w$ and $s$, given the edge profile.

The functional forms used for the edge shape are the following. (Fine details about the functional form describing the edge profile are not crucial.)\\
Panel (a) in Figure~\ref{fig:results}:

\begin{equation}
	g(-y)=\frac{0.05}{|y/w+0.4|^2+0.05}\theta(-y/w+0.5)\theta(y/w+0.5).
\end{equation}
Panel (c) in Figure~\ref{fig:results}:

\begin{equation}
	g(-y)=e^{-(y/w-0.45)^2/(2*0.2^2)}\theta(-y/w+0.5)\theta(y/w+0.5).
\end{equation}
The upper edge (pink in Figures~\ref{fig:wide} and \ref{fig:density}) is simply given by $g(y)$.

Let us start from the first row, where the same edge shape as in the main text is considered. In panel (b), the orange curve is the one presented in Section~\ref{sec:results}, with a high coherence length ($\xi/w=0.85$) and the prominent presence of CAR ($s=0.2$). It is used here as a reference plot.

The black curve shows the opposite regime, where CAR is almost missing ($s=0.7$). Due to $\xi/w\ll1$, one falls back into the Dynes and Fulton description, with the interference pattern approaching the one of a supercurrent density $g(y)^2+g(-y)^2$. This tends to give rise to a standard Fraunhofer-like pattern, with more minima. If $s$ is decreased, LAR is also suppressed, and the entire pattern is lowered.

Increasing the coherence length, the possibility of a nonlocal propagation of the two electrons is enhanced, but it is not sufficient to get a clearly visible $2\phi_0$ periodicity. A LAR-dominated scenario (a weak suppression $s\sim1$), despite high coherence lengths, still leads to Fraunhofer-like behavior with more minima and a slower decay (light blue curve, with $\xi/w=0.85$ and $s=0.7$). This pinpoints the additional demand for a prominent presence of CAR (small $s$, at least $s<1/2$).

The second row allows us to discuss the importance of the overlap of edge states, which is quite small in panel (c). Tuning the parameters as in the black and light blue curves gives a result similar to panel (b). This is expected to be the case, since it has already been commented they are not in the appropriate parameter regime to appreciate the non-local transport significantly. Hence, a more or less pronounced overlap becomes irrelevant. However, using the optimal parameters (orange curve, with $\xi/w=0.85$ and $s=0.2$), the periodicity just starts to approach $2\phi_0$,  but the minima are shallow. This shows the need for highly extended states to see the $2\phi_0$ periodicity.

\section{Conclusions} \label{sec:conclusions}
In this work, a way of computing the supercurrent across a helical Josephson junction that generalizes the previous theoretical approaches by assuming spatially extended edge states has been provided. Strongly localized edge states give rise to a pattern with no decay and a period $\phi_0$ or $2\phi_0$ if Cooper pair splitting over the edges is allowed. Including a finite extent of the edge states in the model gives rise to wider possibilities. A heuristic expression that allows for a simple and intuitive calculation of the Josephson current as a function of the magnetic flux through the junction has been presented. Such expression comes from the generalization to two coordinates of the Dynes and Fulton one, which assumes the electrons within a CP follow the same path. Indeed, it has been argued how, as a consequence of their spatial extension, the edge states can host different trajectories for the two electrons. Some limiting cases have been discussed, showing that the new approach correctly captures the already studied regimes.

The Dynes and Fulton hypothesis of sinusoidal current--phase relation, which holds in the tunneling regime between the superconductors and their link, is maintained by the new approach. A further assumption is that the two edge states have a symmetric profile. On the other hand, the specific functional form describing the edge profile is not crucial. The role played by LAR and CAR processes in determining the interference pattern has been analyzed, together with the importance of the edge states' broadening and of the superconducting coherence length, which represents the size of the Cooper pair. The periodicity of the resultant pattern may vary from $\phi_0$ to $2\phi_0$, depending on the dominating process. 
In particular, the cause for the doubled periodicity has been identified with the non-local transport arrangement. In our case, such non-locality is allowed by the extent of the edges. More specifically, the predicted effects are relevant when the two electrons within a pair can separately explore the two edges and the latter are widely broadened through the junction.

This proposal can help in developing a more realistic description of experimentally realized systems and opens up further generalizations and refinements, such as a justification at a microscopic level of the phenomenological parameters involved.

\vspace{6pt} 



\authorcontributions{Conceptualization, A.C., N.T.Z., and B.T.; Investigation, L.V. and A.C.; Validation, L.V., A.C., N.T.Z., F.S.B., M.S., and B.T.; Writing---original draft, L.V.; Writing---review \& editing, L.V., A.C., N.T.Z., F.S.B., M.S., and B.T. All authors have read and agreed to the published version of the manuscript}

\funding{This work was supported by the ``Dipartimento di Eccellenza MIUR 2018–2022'' and the funding of the European Union--NextGenerationEU through the ``Understanding even--odd criticality’’ project, in the framework of the Curiosity Driven Grant 2021 of the University of Genova. This work was further supported by the W{\"u}rzburg-Dresden Cluster of Excellence ct.qmat, EXC2147, project-id 390858490, and the DFG (SFB 1170). We also thank the Bavarian Ministry of Economic Affairs, Regional Development, and Energy for financial support within the High-Tech Agenda Project “Bausteine f{\"u}r das Quanten Computing auf Basis topologischer Materialen”. The work of F.S.B. was partially supported by the Spanish AEI through project PID2020-114252GB-I00 (SPIRIT),   the Basque Government through grant IT-1591-22, and IKUR strategy program.
F.S.B. acknowledges the A.~v.~Humboldt Foundation for funding and Prof. Trauzettel for the kind hospitality during his stay at W\"urzburg University.}

\institutionalreview{Not applicable.}

\informedconsent{Not applicable.}

\dataavailability{Not applicable.} 


\conflictsofinterest{The authors declare no conflict of interest.} 


\abbreviations{Abbreviations}{
The following abbreviations are used in this manuscript:\\

\noindent 
\begin{tabular}{@{}ll}
JJ & Josephson junction\\
2DTI & two-dimensional topological insulator\\
CP & Cooper pair\\
SQUID & superconducting quantum interference device\\
CAR & crossed Andreev reflection\\
QSH & quantum spin Hall\\
LAR & local Andreev reflection
\end{tabular}
}

\begin{adjustwidth}{-\extralength}{0cm}

\reftitle{References}

\PublishersNote{}
\end{adjustwidth}
\end{document}